\documentstyle[preprint,aps,eqsecnum]{revtex}
\tighten

 \input amssym.def   
 \input amssym.tex

\font\twelvemsb=msbm10 at 12pt
\font\ninemsb=msbm7 at 9pt
\font\sixmsb=msbm5 at 6pt
\newfam\Msbfam
\textfont\Msbfam=\twelvemsb
\scriptfont\Msbfam=\ninemsb
\scriptscriptfont\Msbfam=\sixmsb

\def\half{{\textstyle{1\over2}}}

\def\beq{\begin{equation}}
\def\eeq{\end{equation}}
\def\bi{\begin{itemize}}
\def\ei{\end{itemize}}
\def\beqar{\begin{eqnarray}}
\def\eeqar{\end{eqnarray}}

\newcommand{\Ee}{\mbox{$\cal E\;$}}

\newcommand{\Pp}{\mbox{$\cal P\;$}}
\newcommand{\bcP}{\mbox{\boldmath$\cal P$}}

\newcommand{\bW}{\mbox{\boldmath$\omega$}}
\newcommand{\balpha}{\mbox{\boldmath$\alpha$}}

\newcommand{\r}{\mbox{\bf r}}

\newcommand{\B}{\mbox{\bf B}}

\newcommand{\x}{\mbox{\bf x}}

\newcommand{\bm}{\mbox{\bf m}}
\newcommand{\bu}{\mbox{\bf u}}
\newcommand{\bv}{\mbox{\bf v}}

\newcommand{\bS}{\mbox{\bf S}}
\newcommand{\G}{\mbox{\bf G}}

\newcommand{\bP}{\mbox{\bf P}}
\newcommand{\bp}{\mbox{\bf p}}
\newcommand{\bj}{\mbox{\bf j}}

\newcommand{\drh}{\dot{\rho}}
\newcommand{\dbv}{\dot{\bv}}
\newcommand{\dth}{\dot{\theta}}
\newcommand{\dbet}{\dot{\beta}}
\newcommand{\dpsi}{\dot{\psi}}

\newcommand{\rmd}{{\rm d\null}}

\newcommand{\fract}[2]{{\textstyle\frac{#1}{#2}}}
\def\boldnab{\mbox{\boldmath$\nabla$}}

\def\lra{\mathop{\hbox to .5in{\rightarrowfill}}}

\let\varkappa\kappa

\skip\footins = 14pt 
\begin{document}


\title{Supersymmetric Fluid Mechanics}

\author{R. Jackiw\footnotemark[1]}

\footnotetext[1] {\baselineskip=12pt This work is supported
in part by funds provided by  the U.S.~Department of Energy
(D.O.E.) under contract
\#DE-FC02-94ER40818.  \qquad MIT-CTP-2943 \hfill
 hep-th/004083
}

\address{Center for Theoretical Physics\\ Massachusetts
Institute of Technology\\ Cambridge, MA ~02139--4307,
USA\\[2ex]
\rm and\vspace{-3ex}}

\author{A.P. Polychronakos}

\address{Institute of Theoretical Physics,
Uppsala University\\
S-75108 Uppsala,  Sweden\\
\emph{and} Physics Department, University of Ioannina\\
45110 Ioannina, Greece}

\maketitle\vspace*{-1pc}

\begin{abstract}%
When anticommuting Grassmann variables are introduced
into a fluid dynamical model with irrotational velocity and
no vorticity, the velocity acquires a nonvanishing curl and
the resultant vorticity is described by Gaussian potentials
formed from the Grassmann variables.  Upon adding a
further specific interaction with the Grassmann degrees of
freedom, the model becomes supersymmetric.

\end{abstract}

\section{Introduction} 

An isentropic fluid is described by a matter
density field $\rho$ and a velocity field $\bv$.  These
satisfy the continuity equation, which involves the current
$\bj=\bv \rho$,
\beq
\drh + \boldnab \cdot (\rho \bv)=0
\label{eq:1.1}
\eeq
and the force equation
\beq
\dbv + \bv \cdot \boldnab 
\bv= -\frac{1}{\rho} \boldnab P
\label{eq:1.2}
\eeq
where $P$ is the pressure.  (Over-dot denotes differentiation
with respect to time.) We show that it is possible to
supplement the
$(\rho,
\bv)$ bosonic/commuting variables with
Grassmann/anticommuting variables $\psi$ such that the
entire system exhibits a centrally extended
supersymmetry.  Moreover, when the bosonic system is
irrotational, so that its vorticity vanishes,
$\omega_{ij} \equiv
\partial_iv^j-\partial_jv^i=0$, and the velocity is the gradient
of a velocity potential
$\bv=\boldnab \theta$, the Grassmann variables
give rise to nonvanishing vorticity and provide the Gaussian
potentials in a Clebsch representation for the total velocity
(see below).

The specific system that we analyze devolves from the
dynamics for a membrane (a 2-dimensional extended object),
which propagates in $(3+1)$ dimensional space-time.  The
emergent fluid propagates in two spatial
dimensions.  When the membrane involves just bosonic
variables, the fluid is irrotational~\cite{ref:1,ref:2}.  Our
supersymmetric fluid is derived by a similar construction,
starting from a supermembrane~\cite{ref:3}.

In the remainder of this section, we review the
action/Hamiltonian formulation for the system
(\ref{eq:1.1})--(\ref{eq:1.2}).  In the next section, we
directly present the supersymmetric model.  Section III is
devoted to a derivation of this supersymmetric fluid from a
supermembrane, while concluding remarks comprise the
last Section IV.

For isentropic fluids, the pressure $P$ is a function only of
the density, and the right side of (\ref{eq:1.2}) may also be
written as $-\boldnab V'(\rho)$, where $V'$ is the
enthalpy,
$V''(\rho) = \frac{1}{\rho}P'(\rho)$, and $\sqrt{P'}$ is the
sound speed (dash denotes differentiation with respect to
argument).  Moreover, equations  (\ref{eq:1.1}) and
(\ref{eq:1.2}) can be obtained by (Poisson) bracketing with
the Hamiltonian
\beq
H=\int \rmd r \Big(\half \rho v^2
+V(\rho) \Big) 
\label{eq:1.3}
\eeq
\vspace*{-2.5pc}

\begin{mathletters}\label{eq:1.4}%
\begin{eqnarray}
\drh &=& \{ H, \rho \} \label{eq:1.4a} \\[1ex]
\dbv &=& \{ H, \bv \} \label{eq:1.4b}
\end{eqnarray}
\end{mathletters}%
provided the nonvanishing brackets of the fundamental
variables $(\rho, \bv)$ are taken to be
\begin{mathletters}\label{eq:1.5}%
\begin{eqnarray}
\{ v^i(\r), \rho (\r') \}&=& \partial_i \delta(\r-\r')
\label{eq:1.5a} \\[1ex]
\{ v^i(\r), v^j (\r') \}&=&
-\frac{\omega_{ij}(\r)}{\rho(\r)} \delta(\r-\r')\ .
\label{eq:1.5b}
\end{eqnarray}
\end{mathletters}%
(The fields in the brackets are at equal times, hence the time
argument is suppressed.)  An equivalent, more transparent
version of the algebra (\ref{eq:1.5}) is satisfied by the field
momentum density, which also coincides with the current
$\bj$.
\beq
\Pp = \rho \bv 
\label{eq:1.6}
\eeq
As a consequence of (\ref{eq:1.5}) we have
\begin{mathletters}
\begin{eqnarray}
\{ \Pp^i(\r), \rho (\r') \}&=& \rho(\r) \partial_i
\delta(\r-\r')
\label{eq:1.7a} \\[1ex]
\{ \Pp^i(\r), \Pp^j (\r') \}&=&
\Pp^j(\r) \partial_i  \delta(\r-\r') + \Pp^i(\r') \partial_j
\delta(\r-\r')\ .
\label{eq:1.7b}
\end{eqnarray}
\label{eq:1.7}%
\end{mathletters}%
This is the familiar algebra of momentum densities. The
Jacobi identity is satisfied by (\ref{eq:1.5}) and (\ref{eq:1.7}).
The above holds in any dimension\cite{ref:4new}.

One naturally asks whether there is a canonical 1-form that
leads to the symplectic structure (\ref{eq:1.5}),
(\ref{eq:1.7}); that is, one seeks a Lagrangian whose
canonical variables can be used to derive  (\ref{eq:1.5}) and
(\ref{eq:1.7}) from canonical brackets.  When the velocity is
irrotational, the vorticity vanishes, $\bv$ can be written as
$\boldnab\theta$, and (\ref{eq:1.5}) is satisfied by
postulating that
\beq
\{\theta(\r), \rho(\r') \} = \delta(\r-\r')
\label{eq:1.8}
\eeq
that is, the velocity potential is conjugate to the density,
so that the Lagrangian can be taken as
\beq
L\Bigr|_{\rm irrotational} = \int \rmd r\, \theta \drh -H
\label{eq:1.9}
\eeq
where $H$ is given by (\ref{eq:1.3}) with
$\bv=\boldnab\theta$.

With nonvanishing vorticity, the canonical formulation is
more indirect.  One writes the velocity in a Clebsch
decomposition, which in two and three spatial dimensions
reads
\beq
\bv = \boldnab \theta + \alpha \boldnab \beta\ .
\label{eq:1.10}
\eeq
Then
\beq
L=-\int \rmd r\,  \rho (\dth + \alpha \dbet) -H\ .
\label{eq:1.11}
\eeq
Here $\alpha$ and $\beta$ are the ``Gauss potentials'', and
from (\ref{eq:1.11}) is seen that $\{ \theta, \rho\}$ as well
as $\{ \beta, \rho\alpha \}$ are canonically conjugate.  It
then follows that $\bv$, given by (\ref{eq:1.10}), and $\rho$ 
satisfy  (\ref{eq:1.5}).\footnote{Some more observations on
the Clebsch decomposition of the vector field $\bv$:  In
three dimensions, Eq.~(\ref{eq:1.10}) involves the same
number of functions on the left and right sides of the
equality: three.  Nevertheless the Gauss potentials are not
uniquely determined by $\bv$.  The following is the reason
why a canonical formulation of  (\ref{eq:1.5}) requires using
the Clebsch decomposition  (\ref{eq:1.10}).  Although the
algebra  (\ref{eq:1.5}) is consistent in that the Jacobi
identity is satisfied, it is degenerate in that the kinematic
helicity $h$
$$
h \equiv \half \int \rmd^3 r\,  \bv \cdot (\boldnab \times
\bv) =
\half \int \rmd^3 r\,   \bv \cdot \bW
$$
($\omega^i=\half \epsilon^{ijk}\omega_{jk}$) has vanishing
bracket with $\rho$ and $\bv$.  (Note that $h$ is just the
Abelian Chern-Simons term of $\bv$.)  Consequently, a
canonical formulation requires eliminating the kernel of the
algebra, that is, neutralizing $h$.  This is achieved by the
Clebsch decomposition: $\bv=\boldnab \theta+\alpha
\boldnab\beta$,
$\bW=\boldnab\alpha\times\boldnab\beta$, $\bv
\cdot \bW = \boldnab\theta \cdot
(\boldnab\alpha\times\boldnab\beta) = \boldnab \cdot
(\theta \boldnab \alpha \times \boldnab \beta)$.  Thus in
the Clebsch parameterization the helicity is given by a
surface integral $h = \half \int  \rmd \bS \cdot
\theta(\boldnab\alpha\times\boldnab\beta)$ --- it
possesses no bulk contribution, and the obstruction to a
canonical realization of (\ref{eq:1.5}) is removed.

In two spatial dimensions, the Clebsch parameterization is
redundant, involving three functions to express the two
velocity components.  Moreover, the kernel of (\ref{eq:1.5}) 
in two dimensions comprises an infinite number of quantities
$$
k_n = \int \rmd^2 r\,  \rho \Big(\frac{\omega}{\rho} \Big)^n
$$
for which the Clebsch parameterization offers no
simplification.  (Here $\omega$ is the two-dimensional
vorticity $\omega_{ij}=\epsilon_{ij}\omega$.) Nevertheless, a
canonical formulation in two dimensions also uses Clebsch
variables to obtain an even-dimensional phase space.}

\section{Supersymmetric Fluid Mechanics}

\subsection{The Model}

The bosonic fluid model in two spatial dimensions, which
descends from a bosonic Nambu-Goto action, is supplemented
by Grassmann variables $\psi_a$ that are Majorana 
spinors (real, two-component: $\psi_a^* = \psi_a$, $a=1,2$). 
The Lagrange density reads
\beq
{\cal L} = -\rho (\dth - \fract12 \psi \dpsi) - \half \rho
(\boldnab\theta - \fract12 \psi \boldnab \psi)^2 -
\frac{\lambda}{\rho}  -\frac{\sqrt{2\lambda}}{2} \, \psi
\balpha \cdot \boldnab \psi\ .
\label{eq:2.1}
\eeq
Here $\alpha^i$ are two ($i=1,2$), $2 \times 2$, real
symmetric Dirac ``alpha'' matrices; in terms of Pauli matrices
we can take $\alpha^1=\sigma^1$, $\alpha^2=\sigma^3$. 
Note that the matrices satisfy the following relations, which
are needed to verify subsequent formulas
\begin{eqnarray}
\epsilon_{ab} \alpha^i_{bc} &=& \epsilon^{ij} \alpha^j_{ac}
\nonumber \\[1ex]
\alpha^i_{ab} \alpha^j_{bc} &=& \delta^{ij} \delta_{ac}
-\epsilon^{ij} \epsilon_{ac}
\nonumber \\[1ex]
\alpha^i_{ab} \alpha^i_{cd} &=& \delta_{ac} \delta_{bd}
- \delta_{ab} \delta_{cd} + \delta_{ad} \delta_{bc}
\label{eq:2.2}
\end{eqnarray}
$\epsilon_{ab}$ is the $2 \times 2$ antisymmetric matrix
$\epsilon \equiv i \sigma^2$.  In (\ref{eq:2.1}) $\lambda$ is
a coupling strength, taken positive.  The density-dependent
potential
$V(\rho)={\lambda}/{\rho}$ corresponds to a negative
pressure $P=-2\lambda/\rho$ and to sound velocity
$\sqrt{2\lambda}/\rho$.  These describe the ``Chaplygin
gas''.  The Grassmann term enters with coupling
$\sqrt{2\lambda}$, so chosen to ensure supersymmetry (see
below).  It is evident that the velocity should be defined as 
\beq
\bv = \boldnab \theta - \fract12 \psi \boldnab \psi\ .
\label{eq:2.3}
\eeq
The Grassmann variables directly give rise to a Clebsch
formula for $\bv$, and provide the Gauss potentials.  The
two-dimensional vorticity reads
$\omega=\epsilon^{ij}\partial_i v^j = -\fract12
\epsilon^{ij}\partial_i \psi \partial_j \psi = -\fract12
\boldnab \psi \times \boldnab \psi$. The variables 
$\{\theta, \rho\}$ remain a canonical pair, while the
canonical 1-form in (\ref{eq:2.1}) indicates that the
canonically independent Grassmann variables are
$\sqrt{\rho} \, \psi$ so that the antibracket of the $\psi$'s is
\beq
\{ \psi_a (\r), \psi_b (\r')\} = -\frac{\delta_{ab}}{\rho(\r)}
\delta(\r - \r')\ .
\label{eq:2.4}
\eeq
One verifies that the algebra (\ref{eq:1.5}) or (\ref{eq:1.6})
is satisfied, and further, one has
\begin{mathletters}\label{eq:2.5}%
\begin{eqnarray}
\{ \theta (\r), \psi(\r)\} &=& -\frac{1}{2\rho(\r)}
\psi (\r) \delta (\r-\r') \label{eq:2.5a} \\[1ex]
\{ \bv (\r), \psi (\r')\}
&=& 
-\frac{\boldnab \psi(\r)}{\rho(\r)}
 \delta(\r-\r') \label{eq:2.5b} \\[2ex]
\{ \bcP (\r), \psi(\r') \} &=& -\boldnab \psi(\r)
\delta(\r-\r')\ .
\label{eq:2.5c}
\end{eqnarray}
\end{mathletters}%
\newpage

The equations of motion read
\begin{mathletters}\label{eq:2.6}%
\begin{eqnarray}
\drh + \boldnab \cdot (\rho\bv) &=& 0 \label{eq:2.6a}
\\[1ex] 
\dth + \bv \cdot \boldnab \theta
&=& \half v^2 + \frac{\lambda}{\rho^2}+
\frac{\sqrt{2\lambda}}{2 \rho}\, \psi \balpha \cdot
\boldnab \psi \label{eq:2.6b} \\[1ex] 
\dpsi + \bv \cdot \boldnab \psi 
&=& \frac{\sqrt{2\lambda}}{\rho} \, \balpha
\cdot \boldnab \psi \label{eq:2.6c}
\end{eqnarray}
and together with (\ref{eq:2.3}) they imply
\beq
\dbv + \bv \cdot \boldnab \bv = 
\boldnab \frac{\lambda}{\rho^2} +
\frac{\sqrt{2\lambda}}{\rho}\, \boldnab \psi \balpha \cdot
\boldnab \psi\ .
\label{eq:2.6d}
\eeq
\end{mathletters}%
All these equations may be obtained by bracketing with the
Hamiltonian
\beq
H=\int \rmd^2 r\,  \Bigl(\half \rho v^2 + \frac{\lambda}{\rho}
+\frac{\sqrt{2\lambda}}{2}\, \psi \balpha \cdot \boldnab
\psi \Bigr)
\label{eq:2.7}
\eeq
when (\ref{eq:1.5}), (\ref{eq:1.8}) and  (\ref{eq:2.5}) are
used.

We record the components of the energy-momentum tensor,
and the continuity equations they satisfy.  The energy
density $\Ee=T^{oo}$, given by
\beq
T^{oo} = \half \rho v^2 +
\frac{\lambda}{\rho}+\frac{\sqrt{2\lambda}}{2}\, \psi
\balpha \cdot  \boldnab \psi 
\label{eq:2.8}
\eeq
satisfies a continuity equation with the energy flux $T^{oj}$.
\begin{mathletters}\label{eq:2.9}
\beq
\dot{T}^{oo} + \partial_j T^{oj}= 0
\label{eq:2.9a} 
\eeq
\beq
T^{oj}
= \half \rho v^2 v^j - \frac{\lambda v^j}{\rho} + 
\frac{\sqrt{2\lambda}}{2}\, \psi \alpha^j \bv \cdot 
\boldnab \psi - \frac{\lambda}{\rho} \psi \partial_j \psi
+  \frac{\lambda}{\rho} \epsilon^{jk} \psi \epsilon
\partial_k \psi 
\label{eq:2.9b} 
\eeq
\end{mathletters}%
This ensures that the total energy, that is, the
Hamiltonian, is time-independent.  Conservation of the total
momentum
\beq
\bP = \int \rmd^2 r\,  \bcP
\label{eq:2.10}
\eeq
follows from the continuity equation satisfied by the
momentum density ${\cal P}^i=T^{io}$
\begin{mathletters}
\beq
\dot T^{io} + \partial_j T^{ij}= 0
\label{eq:2.11a} 
\eeq
\beq
T^{ij}= \rho v^i v^j - \delta^{ij} \Bigl(\frac{2\lambda}{\rho} 
+ \frac{\sqrt{2\lambda}}{2}\,  \psi \balpha \cdot \boldnab
\psi \Bigr) +  \frac{\sqrt{2\lambda}}{2}\, \psi \alpha^j
\partial_i \psi
\label{eq:2.11b} 
\eeq
\label{eq:2.11}%
\end{mathletters}%
but the momentum flux $T^{ij}$, that is, the stress tensor, is not
symmetric in its spatial indices, owing to the presence of spin
in the problem.  However, rotational symmetry makes it
possible to effect an ``improvement'', which modifies the
momentum density by a total derivative term, leaving the
integrated total momentum unchanged (provided surface
terms can be ignored) and rendering the stress tensor
symmetric.  The improved quantities are
\beq
\hskip-1.5in {\cal P}^i_I = T^{io}_I = \rho v^i + \fract18
\epsilon^{ij}
\partial_j (\rho \psi \epsilon \psi)
\label{eq:2.12new} 
\eeq
\vspace*{-1pc}
\begin{mathletters}\label{eq:2.13}%
\beq
\dot{T}^{io}_I + \partial_j T^{ij}_I = 0
\label{eq:2.13a} 
\eeq
\begin{eqnarray}
 T^{ij}_I
&=& \rho v^i v^j -\delta^{ij} \Bigl(\frac{2 \lambda}{\rho} + 
\frac{\sqrt{2\lambda}}{2} \, \psi \balpha \cdot \boldnab
\psi \Bigr) + \frac{\sqrt{2\lambda}}{4} \, \Big( \psi 
\alpha^i \partial_j \psi + \psi \alpha^j \partial_i \psi  \Big)
\nonumber \\ 
&& \quad {}-
\fract18 \partial_k \Big[ (\epsilon^{ki} v^j + \epsilon^{kj}
v^i) \rho \psi \epsilon\psi\Big] \ .
\label{eq:2.13b} 
\end{eqnarray}
\end{mathletters}%
It immediately follows that the angular momentum
\beq
M=\int \rmd^2 r\,  \epsilon^{ij} r^i {\cal P}^j_I
= \int \rmd^2 r\, \rho \epsilon^{ij} r^i v^j + \fract14\int \rmd
r\, \rho\psi\epsilon\psi 
\label{eq:2.15} 
\eeq
is conserved. The first term is clearly the orbital part
(which still receives a Grassmann contribution through $\bv$),
whereas the second, coming from the improvement, is the spin
part. Indeed, since
$\fract i2\epsilon=\fract12 \sigma^2 \equiv \Sigma$, we
recognize this as the spin matrix in (2+1)~dimensions. The
extra term in the improved momentum density $\fract18
\epsilon^{ij}\partial_j (\rho\psi\epsilon\psi)$ can then be
readily interpreted as an additional localized momentum
density, generated by the nonhomogeneity of the spin density.
This is analogous to the magnetostatics formula giving the
localized current density~$\bj_m$ in a magnet in terms of its
magnetization $\bm$: $\bj_m = \boldnab\times\bm$. All in
all, we are describing a fluid with spin.

Also the total number 
\beq
N=\int \rmd^2 r\,  \rho
\label{eq:2.15new} 
\eeq
is conserved
by virtue of the continuity equation (\ref{eq:2.6a}) satisfied
by $\rho$.  Finally, the theory is Galileo invariant, as is seen
from the conservation of the Galileo boost,
\beq
\B = t \bP-\int \rmd^2 r\,  \r \rho
\label{eq:2.16} 
\eeq
which follows from (\ref{eq:2.6a}) and (\ref{eq:2.10}).  The
generators $H, \bP, M, \B$ and $N$ close on the (extended)
Galileo group.  [The theory is not Lorentz invariant in
$(2+1)$-dimensional space-time, hence the energy flux
$T^{oi}$ does not coincide with the momentum density,
improved or not.]

We observe that $\rho$ can be eliminated from
(\ref{eq:2.1}) so that ${\cal L}$ involves only $\theta$ and
$\psi$.  From (\ref{eq:2.6b}) and (\ref{eq:2.6c}) it follows
that
\beq
\rho=\Big(\dth - \fract12 \psi \dpsi + \half
v^2\Big)^{-\half}\ .
\label{eq:2.17} 
\eeq
Substituting into (\ref{eq:2.1}) leaves
\beq
{\cal L} = - \sqrt{2\lambda} \, \Big\{\sqrt{ 2\dth - \psi
\dpsi +  (\boldnab \theta - \half \psi \boldnab \psi)^2}
+ \fract12 \psi \balpha \cdot \boldnab \psi \Big\}\ .
\label{eq:2.18} 
\eeq
Note that the coupling strength has disappeared from the
dynamical equations, remaining only as a normalization factor
for the Lagrangian.  Consequently the above elimination of
$\rho$ cannot be carried out in the free case, $\lambda=0$.

\subsection{Supersymmetry}

The theory also possesses supersymmetry.  This
can be established, first of all, by verifying that the
following two-component dynamics-dependent supercharges
are time-independent  Grassmann quantities.
\beq
Q_a =  \int \rmd^2 r\,  \Big[ \rho \bv \cdot (\balpha_{ab}
\psi_b) + \sqrt{2\lambda} \psi_a \Big]\ .
\label{eq:2.19} 
\eeq
Taking a time derivative and using the evolution equations
(\ref{eq:2.6}) establishes that $\dot{Q}_a=0$.

Next, the transformation rule for the dynamical variables is
found by considering the Grassmann charge contracted with
a constant Grassmann parameter $\eta^a$, giving a bosonic
symmetry generator $Q=\eta^aQ_a$.  Using the canonical
brackets one verifies the field transformation rules
\begin{mathletters}
\begin{eqnarray}
\delta \rho = \{ Q, \rho\} &=& -\boldnab \cdot
\rho(\eta\balpha\psi)
\label{eq:2.20a} \\[1ex]
 \delta \theta = \{ Q, \theta\} &=&
-\half(\eta\balpha \psi) \cdot \boldnab \theta - \fract14
 (\eta\balpha\psi) \cdot \psi \boldnab\psi +
\frac{\sqrt{2\lambda}}{2\rho}\, \eta  \psi
\label{eq:2.20b} \\[1ex] 
 \delta \psi = \{ Q, \psi\} &=&
-(\eta\balpha \psi) \cdot \boldnab \psi -
\bv \cdot \balpha \eta - \frac{\sqrt{2\lambda}}{\rho}\,\eta
\label{eq:2.20c} \\[1ex] 
 \delta \bv = \{ Q, \bv\} &=&
-(\eta\balpha \psi) \cdot \boldnab \bv +
\frac{\sqrt{2\lambda}}{\rho}\,
\eta \boldnab \psi\ .
\label{eq:2.20d}
\end{eqnarray}
\label{eq:2.20}%
\end{mathletters}%
Supersymmetry is
reestablished by determining the variation of the action $\int
\rmd t\,  L$, consequent to the above field variations:  the
action is invariant.  One then reconstructs the supercharges
(\ref{eq:2.19}) by Noether's theorem. 
Finally, upon computing the bracket of two supercharges,
one finds
\beq
\{ \eta^a_1 Q_a, \eta^b_2 Q_b\} = 2  (\eta_1 \eta_2) H
\label{eq:2.21} 
\eeq
which again confirms that the charges are time-independent:
\beq
\{ H, Q_a\} = 0\ .
\label{eq:2.22} 
\eeq

Additionally a further, kinematical, supersymmetry can be
identified.  According to the equations of motion the following
two supercharges are also time-independent:
\beq
\tilde{Q}_a = \int \rmd^2 r\, \rho \psi_a\ .
\label{eq:2.23} 
\eeq
$\tilde{Q}=\tilde{\eta}^a \tilde{Q}_a$ effects a shift of the
Grassmann field:
\begin{mathletters}\label{eq:2.24} 
\begin{eqnarray}
\tilde{\delta} \rho = \{ \tilde{Q}, \rho\} &=& 0
\label{eq:2.24a} \\[1ex]
\tilde{\delta} \theta = \{ \tilde{Q}, \theta\} &=&
-\half( \tilde{\eta}\psi) 
\label{eq:2.24b} \\[1ex] 
\tilde{\delta} \psi = \{ \tilde{Q}, \psi\} &=&
-\tilde{\eta}
\label{eq:2.24c} \\[1ex] 
\tilde{\delta} \bv = \{ \tilde{Q}, \bv\} &=& 0\ .
\label{eq:2.24d}
\end{eqnarray}
\end{mathletters}
This transformation leaves the Lagrangian invariant, and
Noether's theorem reproduces (\ref{eq:2.23}).  The algebra
of these charges closes on the total number $N$.
\beq
\{ \tilde{\eta}_1^a \tilde{Q}_a, \tilde{\eta}_2^b \tilde{Q}_b \} 
=  (\tilde{\eta}_1 \tilde{\eta}_2) N
\label{eq:2.25} 
\eeq
while the algebra with the generators (\ref{eq:2.19}), closes
on the total momentum, together with a central extension,
proportional to volume of space $\Omega = \int \rmd^2 r$
\beq
\{ \tilde{\eta}^a \tilde{Q}_a, \eta^b Q_b \} 
=   (\tilde{\eta}\balpha \eta) \cdot \bP +  \sqrt{2\lambda}\,
(\tilde{\eta}\epsilon \eta) \Omega\ .
\label{eq:2.26} 
\eeq
The supercharges $Q_a, \tilde Q_a$, together with the Galileo
generators ($H$, $\bP$, $M$, and $\B$),  with
$N$ form a superextended Galileo algebra. The
additional, nonvanishing brackets are
\begin{eqnarray}
 \{ M, Q_a\} &=& \fract12 \epsilon^{ab}Q_b
\label{eq:2.27} \\[1ex]
 \{ M, \tilde Q_a\} &=& \fract12 \epsilon^{ab}\tilde Q_b
\label{eq:2.28} \\[1ex] 
 \{\B, Q_a\} &=& \balpha_{ab} \tilde Q_b\ .
\label{eq:2.29}
\end{eqnarray}

\section{Membrane Connection}

The equations for a supersymmetric Chaplygin fluid devolve
from the supermembrane Lagrangian, $L_M$.  We shall
give two different derivations of this result, which make use
of two different parameterizations for the
parameterization-invariant membrane action and give rise,
respectively, to (\ref{eq:2.1}) and (\ref{eq:2.18}).

We work in a light-cone gauge-fixed theory:  The membrane
in 4-dimensional space-time is described by coordinates
$x^\mu$ $(\mu=0,1,2,3)$, which are decomposed into
light-cone components $x^\pm=\frac{1}{\sqrt{2}} (x^0 \pm
x^3)$ and transverse components $x^i$ $\{i=1,2\}$.  These
depend on an evolution parameter $\tau$ and two
space-like parameters $\phi^r$ $\{r=1,2\}$.  Additionally
there are two-component, real Grassmann spinors $\psi$,
which also depend on $\tau$ and $\phi^r$.  In the light-cone
gauge, $x^+$ is identified with $\tau$, $x^-$ is renamed
$\theta$, and the supermembrane Lagrangian is~\cite{ref:3}
\beq
L_M=\int \rmd^2 \phi\,  {\cal L}_M = -\int \rmd^2 \phi\, 
\{\sqrt{G} -
\fract12
\epsilon^{rs}
\partial_r \psi \balpha \partial_s \psi \cdot \x \}
\label{eq:3.1} 
\eeq
where $G=\det G_{\alpha\beta}$;
\begin{eqnarray}
G_{\alpha\beta} &=& \left(
\begin{array}{cr}
G_{oo} &\quad  G_{os} \\[1ex]
G_{ro} & -g_{rs}
\end{array}
\right)
\nonumber  \\[1ex]
&=& \left(
\begin{array}{cr}
2 \partial_\tau \theta-\partial_\tau \x \cdot \partial_\tau\x
- \psi \partial_\tau \psi & \quad u_s \\[1ex] 
u_r & -g_{rs}
\label{eq:3.2}
\end{array}
\right)\\[2ex]
G &=& g\Gamma
\nonumber \\[1ex]
\Gamma &\equiv&  2\partial_\tau \theta-\partial_\tau \x
\cdot \partial_\tau \x -  \psi \partial_\tau \psi
+g^{rs} u_r u_s \nonumber  \\[1ex] 
g_{rs} &\equiv& \partial_r \x \cdot \partial_s \x
\ ,  \quad g=\det g_{rs} \nonumber  \\[1ex] 
u_r &\equiv& \partial_r \theta - \fract12 \psi
\partial_r\psi - \partial_\tau \x \cdot \partial_r \x \ .
\label{eq:3.3}
\end{eqnarray}
Here $\partial_\tau$ signifies differentiation with respect to
the evolution parameter $\tau$, while $\partial_r,
\partial_s$ differentiate with respect to the space-like
parameters $(\phi^r, \phi^s)$, and $g^{rs}$, the inverse of
$g_{rs}$, is used to move the $(r,s)$ indices.  Note that
the dimensionality of the transverse coordinates $x^i$ is the
same as of the parameters $\phi^r$, namely two.

\subsection{First Derivation}

To give our first derivation, we rewrite the Lagrangian in
canonical, first-order form, with the help of canonical
momenta defined by
\begin{mathletters}\label{eq:3.4}
\begin{eqnarray}
\frac{\partial {\cal L}_M}{\partial \partial_\tau \x} &=&
\bp = -\Pi  \partial_\tau \x - \Pi u^r \partial_r \x
\label{eq:3.4a} \\[1ex]
\frac{\partial {\cal L}_M}{\partial \partial_\tau \theta} &=&
\Pi = \sqrt{{g}/{\Gamma}}
\label{eq:3.4b}
\end{eqnarray}
\end{mathletters}%
\begin{eqnarray}
{\cal L}_M&=&\bp \cdot \partial_\tau \x + \Pi \partial_\tau
\theta - \fract12\Pi \psi \partial_\tau\psi + \frac{1}{2\Pi}
(p^2+g) +
\fract12 \epsilon^{rs}
\partial_r \psi \balpha \partial_s \psi \cdot \x 
\nonumber\\[1ex] 
&&\quad {}+ u^r \Big(\partial_r \x \cdot \bp + \Pi \partial_r
\theta -
\fract12 \Pi \psi \partial_r \psi \Big)\ .
\label{eq:3.5} 
\end{eqnarray}
In (\ref{eq:3.5}) $u^r$ serves as a Lagrange multiplier
enforcing a subsidiary condition on the canonical variables. 
The equations that follow from (\ref{eq:3.5}) coincide with
the Euler-Lagrange equations for
(\ref{eq:3.1})--(\ref{eq:3.3}).  The theory still possesses an
invariance against redefining the spatial parameters with a
$\tau$-dependent function of the parameters.  This freedom
may be used to set $u_r$ to zero and fix $\Pi$ at $-1$.  Next
we introduce the hodographic transformation~\cite{ref:4},
whereby independent-dependent variables are
interchanged, namely we view the $\phi^r$ to be
functions of $x^i$.  It then follows that the constraint on
(\ref{eq:3.5}), which with
$\Pi=-1$ reads
\begin{mathletters}
\beq
\partial_r \x \cdot \bp - \partial_r \theta + \fract12 \psi
\partial_r \psi =0
\label{eq:3.6a}%
\eeq
becomes
\beq
\partial_r \x \cdot \Big(\bp -\boldnab \theta +
\fract12 \psi\boldnab \psi \Big) =0\ .
\label{eq:3.6b}%
\eeq
Here $\bp$, $\theta$ and $\psi$ are viewed as functions of
$\x$, renamed $\r$, with respect to which acts the gradient
$\boldnab$.  Also we rename $\bp$ as $\bv$, which
according to (\ref{eq:3.6b}) is
\beq
\bv = \boldnab \theta -
\fract12 \psi\boldnab \psi\ .
\label{eq:3.6c}%
\eeq
\end{mathletters}%

From the chain rule, it follows that
\beq
\partial_\tau = \partial_t + \partial_\tau \x \cdot
\boldnab
\label{eq:3.7}
\eeq
and according to (\ref{eq:3.4a}) (at $\Pi=-1$, $u^r=0$)
$\partial_\tau \x = \bp = \bv$.  Finally, the measure 
transforms according to $\rmd^2 \phi\,  \to
\rmd^2 r\,  \frac{1}{\sqrt{g}}$.  Thus the Lagrangian
for (\ref{eq:3.5}) becomes, after setting $u^r$ to zero and
$\Pi$ to $-1$,
\beq
L_M=\!\int \!\frac{\rmd^2 r}{\sqrt{g}} \Big( v^2 -\dth - \bv
\!\cdot\!
\boldnab \theta+ \fract12 \psi(\dpsi + \bv \cdot \boldnab
\psi) -\half(v^2+g) -\fract12 \epsilon^{rs}\, \psi
\alpha^i\, \partial_j \psi \, \partial_s x^j\,
\partial_r x^i
\Big)\ .
\label{eq:3.8}
\eeq
But $\epsilon^{rs} \partial_s x^j \partial_r x^i = \epsilon^{ij}
\det \partial_r x^i = \epsilon^{ij} \sqrt{g}$.  After
$\sqrt{g}$ is renamed $\sqrt{2\lambda}/\rho$,
(\ref{eq:3.8}) finally reads
\beq
L_M=\Bigl(\frac1{\sqrt{2\lambda}}\Bigr) \int \rmd^2 r\,  \Big(
{-\rho} (\dth -
\fract12 \psi\dpsi) -
\half \rho (\boldnab \theta - \fract12 \psi \boldnab
\psi)^2 - \frac{\lambda}{\rho} - \frac{\sqrt{2\lambda}}{2} 
\psi \balpha \times \boldnab \psi \Big)\ .
\label{eq:3.9}
\eeq
Upon replacing $\psi$ by $\frac{1}{\sqrt{2}} (1-\epsilon)
\psi$, this is seen to reproduce the Lagrange density
(\ref{eq:2.1}), apart from an overall factor.

\subsection{Second Derivation}

For our second derivation, we return to
(\ref{eq:3.1})--(\ref{eq:3.3}) and use the remaining
reparameterization freedom to equate the two $x^i$
variables with the two $\phi^r$ variables, renaming both as
$r^i$~\cite{ref:5}.  Also $\tau$ is renamed as $t$. In
(\ref{eq:3.1})--(\ref{eq:3.3}) $g_{rs} = \delta_{rs}$, and
$\partial_\tau \x=0$, so that (\ref{eq:3.3}) becomes simply
\begin{eqnarray}
G=\Gamma&=&2\dth -  \psi \dpsi + u^2
\label{eq:3.10} \\ [1ex]
 \bu &=& \boldnab \theta - \fract12 \psi \boldnab \psi\ .
\label{eq:3.11}
\end{eqnarray}
Therefore the Nambu-Goto action (\ref{eq:3.1}) reads
\beq
L_M=-\int \rmd^2 r\,  \Bigl\{ \sqrt{2 \dth -  \psi \dpsi +
\bigl(\boldnab \theta -\fract12 \psi \boldnab \psi\bigr)^2}
+ \fract12 \psi \balpha \times \boldnab \psi \Bigr\}\ .
\label{eq:3.12}
\eeq
Again a replacement of $\psi$ by $\frac{1}{\sqrt{2}}
(1-\epsilon) \psi$ demonstrates that the integrand
coincides with the Lagrange density in (\ref{eq:2.18}) (apart
from a normalization factor).

\subsection{Further Consequences of the Supermembrane
Connection}

The supermembrane dynamics is Poincar\'e invariant in
(3+1)-dimensional space-time.  This invariance is hidden
by the choice of light-cone parameterization: only the
light-cone subgroup of the Poincar\'e group is left as a
manifest invariance.  This is just the $(2+1)$ Galileo group
generated by $H$, $\bP$, $M$, $\B$, and $N$.  (The
light-cone subgroup of the Poincar\'e group is isomorphic to
the Galileo group in one lower dimension.)  The Poincar\'e
generators not included in the above list correspond to
Lorentz transformations in the ``$-$'' direction.  We expect
therefore that these generators are ``dynamical'', that is,
hidden and unexpected conserved quantities of our
supersymmetric Chaplygin gas, similar to the situation with
the purely bosonic model~\cite{ref:6}.

One verifies that the following quantities
\begin{eqnarray}
D &=& tH-\int \rmd^2 r\,  \rho \theta 
\label{eq:3.13} \\[1ex]
\G &=& \int \rmd^2 r\,  (\r \Ee - \theta \bcP_I - \fract18
\psi \balpha \balpha \cdot \bcP_I \psi) 
\label{eq:3.14} \\
&=& \int \rmd^2 r\,  (\r \Ee  - \theta \bcP - \fract14
\psi \balpha \balpha \cdot \bcP \psi) 
\label{eq:3.15}
\end{eqnarray}
are time-independent by virtue of the equations of motion
(\ref{eq:2.6}), and they supplement the Galileo generators to
form the full $(3+1)$ Poincar\'e algebra, which becomes the
super-Poincar\'e algebra once the supersymmetry is taken
into account.

\section{Conclusion}
We have shown how fluid dynamics can be extended to
include Grassmann variables, which also enter in a
supersymmetry-preserving interaction.  Since our
construction is based on a supermembrane in
(3+1)-dimensional space-time, the fluid model is
necessarily a planar Chaplygin gas.  It remains to be shown
how this construction could be extended to arbitrary
dimensions and to different interactions. Note that Grassmann
Gauss potentials can be used even in the absence of
supersymmetry. For example, our theory (\ref{eq:2.1}), with
the last term omitted, posseses a conventional, bosonic
Hamiltonian without supersymmetry, while the Grassmann
variables are hidden in $\bv$ and occur only in the canonical
1-form. In a related investigation, conventional fluid
mechanics is generalized, so that it possesses a non-Abelian
gauge symmetry~\cite{ref:7}.

\paragraph*{Note Added:} J.~Hoppe has informed us that some
of the above results were obtained by him in unpublished
research: Karlsruhe preprints KA-THEP-6-93 and 
KA-THEP-9-93 (hep-th/9311059).

\end{document}